\documentclass[mathleft]{an}
\usepackage{graphicx}
\usepackage{times}
\def\e{{\scriptsize$\pm$}} 
\def\kms{km$\,$s$^{-\!1}$} 
\def\vsi{$v\: \sin i$} 
\def\rsun{R$_{\odot}$}

\def\msun{M$_{\odot}$}
\newcommand{\ltsima} {$\; \buildrel < \over \sim \;$} 
\newcommand{\simlt} {\lower.5ex\hbox{\ltsima}} % < over MMM 
\newcommand{\gtsima} {$\; \buildrel > \over \sim \;$} 
\newcommand{\simgt} {\lower.5ex\hbox{\gtsima}} % > over MMM 
\def\t{{\scriptsize$\times$}}

\begin{document}

% The following seven commands are intended for editorial usage and should be ignored by
% the author(s).
\Pagespan{789}{}% Document's page range. 
% If second parameter is left empty, the last page is computed automatically.
\Yearpublication{2009}%
\Yearsubmission{2008}%
\Month{11}%   
\Volume{999}%  
\Issue{88}% 
% \DOI{This.is/not.aDOI}% 

\title{Tidal Interaction in High Mass X-ray Binaries}

\author{K. A. Stoyanov\inst{1}\fnmsep\thanks{Corresponding author:
  \email{kstoyanov@astro.bas.bg}\newline}
%Example 
%for footnote, note the usage of the \texttt{fnmsep}
%command as separator between institute number and footnote mark} 
\and R. K. Zamanov\inst{1}
}
\titlerunning{Tidal interaction - HMXRBs}
\authorrunning{Stoyanov \& Zamanov}
\institute{Institute of Astronomy, Bulgarian Academy of Sciences, 
       72 Tsarighradsko Shousse Blvd., 1784 Sofia, Bulgaria}

\received{25 Nov 2008}
\accepted{05 May 2009}
\publonline{later}


\abstract{%
Our aim is to investigate tidal interaction
in  High-Mass X-ray Binary stars in order to determine in 
which objects the rotation of the mass donors 
is synchronized or pseudosynchronized with 
the orbital motion of the compact companion. 
We calculate the pseudosynchronization period (P$_{ps}$) and compare 
it with the rotational period of the mass donors (P$_{rot}$). 
We find that (1) the Be/X-ray binaries are not synchronized, 
the mass donors rotate faster than the orbital period
and the ratio $P_{ps}/P_{rot}$ is 2 -- 300;
(2) the giant and supergiant systems are close to synchronization
and for them the ratio $P_{ps}/P_{rot}$ is 0.3 -- 2. 
}
\maketitle

\section{Introduction} 
The High-Mass X-ray Binaries (HMXRBs) contain a primary star of spectral type O or B
and a compact object (neutron star or black hole) as a companion. 
The mass donors have a mass  greater then 10~\msun. 
They  are population I objects and are concentrated in the Galactic plane. 
In the majority of HMXRBs the neutron star is detected as an X-ray pulsar.
They are separated in two groups: 
(1) High-Mass Supergiant and Giant 
Systems, where the mass donor is an O-B giant or supergiant.
Accretion is realized by Roche lobe overflow or via a stellar wind or 
a combination of both. 
(2) Be/X-ray binaries, where the mass donor is a main sequence Be star. 
The compact object accretes mainly from the dense circumstellar disk around
the Be star, although the accretion from a polar wind also has some contribution.

The last edition of the catalogue of the galactic HMXRBs contains 
114 objects (Liu,  van Paradijs \& van den Heuvel 2006). 
Among the 68 objects with identified primaries  $\sim60$\% are Be/X-ray binaries,
$\sim$15\%  are classified as giants, and $\sim$22\% are supergiant/X-ray binaries.
The catalogue of HMXRBs in the Magellanic Clouds contains 128 binaries
(Liu,  van Paradijs \& van den Heuvel 2005), where most of the  objects (76\%) 
are known or suspected Be/X-ray binaries. 

As a result of the accretion of matter from the OB primary,
the compact object in these binaries is a
strong X-ray emitter and displays different types of activity
(cf.  Campana et al. 1995; Okazaki \& Negueruela 2001; Ducci et al. 2008
and references therein). 
The rotation of the neutron stars in HMXRBs has been investigated  
in detail with the X-ray satellites (see Bildsten et al. 1997) and their spin period is connected
with the orbital period (Corbet 1984).

Our aim here is to check whether the rotation of the mass donors in HMXRBs is synchronized 
with the orbital motion of the compact object, and how the presence of 
an orbiting neutron star and the tidal force influences the rotation of the mass donor.

\section {Synchronization and pseudosynchronization}

\subsection{Rotation and pseudosynchronization}
In a binary with a circular orbit the rotational period of the primary, P$_{rot}$, 
reaches an equilibrium value at the orbital period, $P_{orb}=P_{rot}$.  
In other words the synchronous rotation (synchronization) means that 
the rotational period is equal to the orbital period. 
In a binary with an eccentric orbit,
the corresponding equilibrium is reached at a value of $P_{rot}$ which is
$less$ than $P_{orb}$, the amount less being a function solely of the orbital
eccentricity $e$. In practice, in a binary with an eccentric orbit the 
tidal force acts to synchronize the rotation of the mass donor with 
the motion of the compact object at the periastron - 
pseudosynchronous rotation (Hall 1986).
To calculate the period of pseudosynchronization, P$_{ps}$, we use (Hut 1981):
\begin{equation}
P_{ps} = \frac{(1+3e^2+\frac{3}{8}e^4)(1-e^2)^\frac{3}{2}}{1+\frac{15}{2}e^2+
\frac{45}{8}e^4+\frac{5}{16}e^6} P_{orb}.
\label{Eq-ps}
\end{equation}
At low eccentricity of the orbit 
$e \rightarrow 0$ and P$_{ps} \approx  P_{orb}$.

To calculate P$_{rot}$ we use
\begin{equation}
P_{rot}=\frac{2\pi R_1 \sin i}{v\: \sin i}, 
\label{Eq-Prot}
\end{equation}
where \vsi\ is the projected rotational velocity 
of the mass donor, and $i$ is the inclination of the orbit to the line of sight. 
The underlying assumption is that the rotational axis of the mass donor is perpendicular to the orbital plane. 

\subsection{Rotation of the mass donors in HMXRBs}

We searched in Liu, van Paradijs \& van den Heuvel (2000) and in the literature and find 13 HMXRBs with well measured 
orbital and stellar parameters. 
The data are collected in Tables~1 and 2. 
These are objects 
for which we were able to find the spectral type of the mass donor, orbital period,
eccentricity of the orbit, inclination ($i$), 
\ and projected rotational velocity of the primary (\vsi). 
The sources of data for each object are given  in Sect.~\ref{indiv}.

Using Eq.~\ref{Eq-ps},  Eq.~\ref{Eq-Prot} 
and the data collected in Table~1, we calculated P$_{ps}$ and P$_{rot}$ for the objects in our sample.
The values are given in Table 1. 

\begin{figure}
 \mbox{} 
 \vspace{8.8cm}  
 \includegraphics{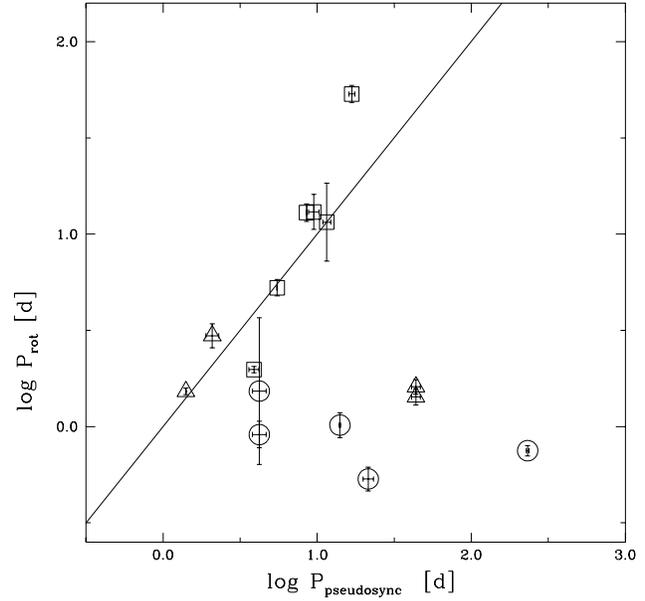}   
 \caption[]{P$_{rot}$ versus P$_{ps}$ on a logarithmic scale. 
      The straight line indicates P$_{ps}=$P$_{rot}$.  
      The circles indicate the Be/X-ray binaries (luminosity class V), the triangles those with 
      giant primaries (luminosity class III), the squares - supergiants (luminosity class I).
      For the systems LSI$+61^0303$  and V725 Tau we found two values for \vsi, so there are two  
       points for each of them. }
    \label{Pps-rot}
\end{figure}

In Fig.~\ref{Pps-rot} we plot P$_{rot}$ versus P$_{ps}$. 
In this figure it is seen that the objects where the mass donors are from spectral class I 
are located close to the line 
P$_{ps} = P_{rot}$ (synchronization/pseudosynchronization), 
while those with mass donors from spectral class V are far away from the equilibrium state.

%---------------------------------------------------------------------
\subsection{Circularization and synchronization time scales}

Following Hurley, Tout \& Pols (2002) and Zahn (1975) the circularization 
timescale for stars with  radiative envelopes can be estimated as:
\begin{equation}
\frac{1}{{\tau}_{circ}}= 
\frac{21}{2} \left( \frac{G M_1}{R_1^3} \right)^\frac{1}{2} 
q_2 \left(1 + q_2 \right)^\frac{11}{6} E_2 \left(\frac{R_1}{a} \right)^\frac{21}{2},  
\end{equation}
where M$_1$ and R$_1$ are the mass and the radius of the primary respectively, 
q$_2$ is the mass ratio M$_2/$M$_1$, and $a$ is the semi-major axis.
$E_2$ is a second-order tidal coefficient,
\begin{equation}
E_2 = 1.592 \times 10^{-9} M_1^{2.84}.  
\end{equation}

The synchronization time scale (Hurley et al. 2002) is given as 
\begin{equation}
\tau_{sync} = K \tau_{circ}, 
\end{equation}
where K  is: 
\begin{equation}
K \approx \frac{0.015}{r_g} \, \frac{1 + q_2}{q_2} \left( \frac{R_1}{a} \right)^2.  
\end{equation}
For the gyration radius of the primary $r_g^2=I/M_1 R_1^2$ (where I is the moment of inertia), 
we adopt $r_g\approx0.16$ for giants, and $r_g\approx0.25$ for main sequence stars
(Claret \& Gimenez, 1989).

In  Table~\ref{t-times} are given 
the adopted stellar parameters, and the calculated  $\tau_{circ}$  and  $\tau_{sync}$.

\subsection{Lifetimes}
 The lifetime of a star on the main sequence can be estimated as (Germany et al. 2009):
\begin{equation}
 \tau_{MS} = 10^{10} (\frac{M_\odot}{M})^{2.5} \; \; {\rm years}.  
\label{Eq-MS}
\end{equation}
For example, a B0V star with a mass $\sim20$~M$_\odot$ spends $\sim5.5$ \t $10^6$~yr on the 
main sequence. Comparing these  lifetimes with $\tau_{sync}$ 
from Table~\ref{t-times}, 
we see that among the Be/X-ray binaries only for LSI+61$^0$303 is
$\tau_{sync} \sim \tau_{MS}$.
This is the only Be/X-ray binary for which we can expect
considerable changes of the rotation  of the primary
during the lifetime of the Be star. 

In comparison, the lifetime of a giant of $\sim20$M$_\odot$ is about 0.1$\tau_{MS}$ 
($\sim 5$ \t $10^5$~yr). The calculated lifetimes are given in Table~\ref{t-times}.
The lifetime of the giant is comparable or longer then 
$\tau_{circ}$ \ and  \  $\tau_{sync}$  \  (see Table~\ref{t-times}) for the giant/supergiant systems
with short orbital periods (P$_{orb}<$20 d). The exceptions are V725~Tau and BP~Cru, for
which $\tau_{sync}$ and $\tau_{circ}$ are longer than the lifetime of the giant
/supergiant stage. 

For the giant/supergiant systems, we ignore the preceding evolution,  
because (1) during the main-sequence stage the tidal interaction  is considerably weaker,
and (2) changes of the orbit at the supernova explosion.

%%%------------------------------------------------------------------------------  
%%%------------------------------------------------------------------------------- 

%%-------------------------------------------------------------------------------
\begin{table*}
  \begin{center}
  \caption{Orbital parameters of HMXRB stars. Given here are as follows:
  name of the object,  orbital period, 
  eccentricity of the orbit, inclination of the orbit to the line of sight, 
  semi-major axis of the orbit, 
  projected rotational velocity (\vsi) of the mass donor, 
  the period of pseudosynchronization (calculated using Eq.\ref{Eq-ps}), 
  rotational period of the mass donor (calculated using Eq.\ref{Eq-Prot}). }
  \begin{tabular}{llcccrrrr}
\hline 
% \medskip
&&  \\
object      &   P$_{orb}$ &  e     &  i   &  a    & \vsi   & P$_{ps}$ & P$_{rot}$ & \\
            &     [d]     &        & [$^o$]       &[\rsun]& [\kms] & [d]      & [d]       & \\
\hline    
&& \\
\multicolumn{2}{c}{{\bf Be/X-ray binaries}}  \\
LSI$+61^0303$ &  26.496   & 0.72$\pm$15   & 30$\pm$20    & 35.45 &  113 & 1.10 - 7.33 & 0.45-2.61 & \\
              & 	  &	          & 70-80        &	 &  360 &	          & 0.77-1.05  & \\
X Per       &    250$\pm$0.6	  & 0.111$\pm$0.018   & 26-33        &  474  &  215$\pm$10 &  228 - 237 &  0.70 - 0.80 & \\
BQ Cam      &    34.25    & 0.31$\pm$0.03   &  $\le10.3$ $\pm$0.09     & 121   &  145 & 19.82 - 23.11    &  0.46 - 0.61 & \\
V635 Cas    &    24.3	  & 0.34   & 40-60        & 95    &  300 &  14.06    &  0.87 - 1.17  & \\
\\
\multicolumn{2}{c}{{\bf Giant systems}}  \\
V725 Tau    &    111	  & 0.47$\pm$0.02   &  28.5        & 23.39 &  254 &  40.93 - 46.62     &   1.43  & \\
            &   	  &	   &	          &       &  225 &	      & 1.61  & \\
LMC X-4        &  1.4084  & 0.006(2)  &  25-29       & 13.3  &  240$\pm$25 &  1.4079 - 1.4083  &   1.46 - 1.59  & \\
Cen~X-3        &  2.0871  & 0.0016   & 70.2$\pm$2.7 & 17.9  &  200$\pm$40 & 2.09   &   2.54 - 3.39  & \\
\\
\multicolumn{2}{c}{{\bf Supergiant systems}} \\
V830 Cen       & 14.365\e0.002 & 0.20$\pm$0.03   & 40-55        & 58.6  & 150  & 10.87 - 12.23   &   6.51 - 16.58  & \\
LSI$+65^0 010$ & 11.5983       & 0.18$\pm$0.05   & 45           & 56    &96\e20&  8.78- 10.25  &   10.36 - 15.74  & \\
Vela~X-1       &    8.9644     & 0.0898\e0.0012&$76^{+5}_{-9}$&50.1&116$\pm$6&  8.54 - 8.56   &  11.59 - 14.29 & \\
SMC X-1        & 3.89229       &$<\!4.10^{-5}$ & 26-30.5 &  25 & 170\e30   &  3.89   &  1.90 - 2.06  & \\
BP Cru         & 41.498\e0.002 & 0.462$\pm$0.014  & 60$\pm$10 & 182  &  50       &  16.03- 17.54  &  48.08 - 58.98  & \\
Cyg X-1	       &  5.5	       & 0.06$\pm$0.01 & 33$\pm$5	&  15.48 & 100 & 5.48 - 5.52 & 4.75 - 5.79 & \\
\hline 
\end{tabular}
  \label{table1}
  \end{center}
\end{table*}

%------------------------------------------------

\begin{table*}
\begin{center}
  \caption{HMXRB star parameters of the components. Given here are  
  the name of the object, the spectral type of the primary,
  mass of the primary, mass of the secondary, radius
  of the primary, its luminosity, 
  synchronization time scale, 
  circularization time scale, the lifetime,
  the action of the tidal force. }
  \begin{tabular}{lllllllllll}
\hline 
% \medskip
&& \\
object   & Sp  &  M$_1$       & M$_2$       & R$_1$       & L$_1$       & $\tau_{sync}$ & $\tau_{circ}$ & lifetime & Tidal \\
         &     &  [M$_\odot$] & [M$_\odot$] & [R$_\odot$] & [L$_\odot$] & [yr]  & [yr]  	 & [yr]     & force \\
\hline
\\
\multicolumn{2}{c}{{\bf Be/X-ray binaries}}  \\
LSI+61$^0$303 & B0Ve         &  20.0 & 4.0 & 6.7\e0.9& 3\t10$^3$   &  3.1\t10$^6$    & 6.8\t10$^7$    &5.6\t10$^6$& pseudosync/spin-down\\
X~Per         & B0V          &  15.5 & 1.4 & 6.5     & 3\t10$^3$   &  6.2\t10$^{17}$ & 1.8\t10$^{21}$ &1.1\t10$^7$& spin-down	\\
BQ~Cam        & O8-9Ve       &  23.0 & 1.4 & 9.0     & 5.5\t10$^3$ &  3.5\t10$^{11}$ & 7.6\t10$^{13}$ &3.9\t10$^6$& spin-down	\\
V635~Cas      & B0.2Ve       &  18.0 & 1.4 & 8.0     & 3\t10$^3$   &  1.4\t10$^{11}$ & 9.5\t10$^{12}$ &7.3\t10$^6$& spin-down	\\
\\
\multicolumn{2}{c}{{\bf Giant systems}}  \\
V725~Tau      & O9.4IIIe     &  23.0 & 1.4 & 15.0        & 2.0\t10$^5$ & 2.8\t10$^{12}$ & 8.0\t10$^{14}$ &4\t$10^5$&spin-down	 \\
LMC X-4       & O8III        &  15.8 & 1.47& 7.8$\pm$0.3 & 2.0\t10$^5$ & 4.5\t10$^2$    & 7.7\t10$^2$    &1\t$10^6$&sync	 \\
Cen~X-3       & O6.5II-IIIae &  20.5 & 1.4 & 12.1$\pm$0.5& 5.0\t10$^5$ & 2.3\t10$^3$    & 4.2\t10$^3$    &5\t$10^5$&sync	 \\
\\
\multicolumn{2}{c}{{\bf Supergiant systems}} \\
V830~Cen      & B2Iae        &  16.0    & 1.4   & 30-60    & 2.5\t10$^5$ & 7.5\t10$^3$ & 1.4\t10$^4$ & 1\t10$^6$ & pseudosync  \\
LSI+65$^0$010 & B1Iae        &  16\e5   & 1.7   & 37\e15   & 2.5\t10$^5$ & 1.3\t10$^4$ & 3.9\t10$^4$ & 1\t10$^6$ & pseudosync/spin-up  \\
Vela~X-1      & B0Ia         &  23.1    & 1.9   & 30.4\e1.6& 2.5\t10$^5$ & 1.0\t10$^4$ & 2.8\t10$^4$ & 3.9\t10$^5$ & sync/spin-up	   \\
SMC X-1       & B0Ib         &  16.7    & 1.05  & 14\e2    & 2.5\t10$^5$ & 3.3\t10$^4$ & 8.2\t10$^4$ & 8.8\t10$^5$ & spin-down   \\		      
BP Cru        & B1.2Ia       &  43.0    & 1.9   & 62.0     &   5\t10$^5$ & 1.8\t10$^6$ & 8.8\t10$^6$ & 8\t10$^4$ & spin-up     \\
Cyg X-1	      & B0Iab        & 40\e10   & 20\e5 & 20-22    & 3-4\t10$^5$ &  $<1$      & $<1$       & 1\t10$^5$ &  sync   \\
\hline

 \end{tabular}
  \label{t-times}
  \end{center}
\end{table*}

\section{Individual objects}
\label{indiv}

\subsection{Be/X-ray binaries}
{\bf LSI$+61^0303$} (V615 Cas, GT0236+610) - 
The system contains a compact object (probably a black hole) 
orbiting \\ around a Be star in a highly eccentric orbit 
(Hutchings \& Crampton 1981; Casares et al. 2005). 
The parameters of the system are not well defined. 
We calculate  $P_{ps}/P_{rot} \approx 2\pm1$.  LSI$+61^0303$ 
is the closest to pseudosynchronization among
the Be/X-ray binaries in our sample.

{\bf X Per} (4U 0352+30) - the stellar parameters are taken from Roche et al. 
(1997), Delgado-Mart{\'{\i}} et al. (2001) and  Lyubimkov et al. (1997). 
The system is non-synchronized with $P_{ps}/P_{rot}\approx 310 \pm 15$.
The tidal force should spin down the rotation of the mass donor,
however it is very weak ($\tau_{sync}\sim 10^{17}$~yr), 
so there should be no changes during the lifetime as a Be/X-ray binary.
It has persistent X-ray emission because the neutron star accretes from
the outer parts of the stellar wind, where there are no changes in 
the density of the material.

{\bf BQ Cam} (V0332+53) -  we calculate a ratio $P_{ps}/P_{rot}\approx 40 \pm 3$. 
The tidal force spins down the rotation of the mass donor. The stellar
parameters are taken from Negueruela et al. (1999). The lack of recent 
X-ray activity is explained by the fact that the dense
regions of the circumstellar disc around the Oe star do not reach 
the orbit of the neutron star.

{\bf V635 Cas} (4U0115+63) - this system is a transient X-ray emitter. We took the stellar 
parameters from Negueruela et al. (2001). The ratio $P_{ps}/P_{rot}\approx 14 \pm 2$ shows that the tidal
force is spinning down the rotation of the mass donor. The disc around the Be star
was modeled as a viscous decretion disc, i.e., a quasi-Keplerian disc held
by the transport of angular momentum via viscous interactions. The outflow
(radial) velocity in such a disc is expected to be strongly subsonic,
in agreement with all the observations of Be stars in general and V635 Cas
in particular. It was shown that such a disc cannot reach a steady
state due to tidal and resonant interaction with the neutron star,
and it is truncated at a radial distance which depends on the value of the viscosity 
(Negueruela \& Okazaki 2001; Reig et al. 2007). \\

\subsection{Systems containing a giant donor}

{\bf V725 Tau} (1A 0535+262)
- the stellar parameters are taken from Clark et al. (1998),
Haigh, Coe \& Fabregat (2004) and Grundstrom et al. (2007a). 
$P_{ps}/P_{rot}\approx 30 \pm 2 $.         
The tidal force spins down the rotation of the mass donor.
The X-ray source A0535+262 was discovered by Ariel V during a 
large Type II outburst in 1975 (Coe et al. 1975; Rosenberg et al. 1975).
Since then the source has been observed to undergo numerous outbursts,
however there were no reported
detections of X-ray outburst activity from 1994 to 2005 (Coe et al. 2006;
Kretschmar et al. 2006). The source reappeared in a Type II outburst in
May/June 2005 and was detected by Swift (Tueller et al. 2005) and RHESSI
(Smith et al. 2005). It was subsequently seen to undergo a Type I outburst in
August 2005 (Kretschmar et al. 2006; Caballero et al. 2007).

In respect to its X-ray behaviour and rotation of the mass donor 
(and ratio $P_{ps}/P_{rot}$) this object is similar to the transient  Be/X-ray binaries.

{\bf LMC~X-4} - 
Kelley et al. (1983) 
discovered the 13.5 s X-ray pulsations of LMC X-4.
The optical light curve shows ellipsoidal variations and a super-orbital 
period of $\sim$30 d (Heemskerk \& van Paradijs 1989). 
The X-ray light curve includes regular 
eclipses as well as a pronounced flux modulation of a factor $\sim$60 
with a period of 30.5 d (Lang et al. 1981). This long-term variation 
is attributed to the precessing accretion disc.
The stellar parameters are taken from van der Meer et al. (2007).
$P_{ps}/P_{rot}\approx 0.92 \pm 0.04 $.
The rotation of the mass donor is synchronized with the orbital
motion. In this respect it is similar to the supergiant systems.

{\bf Cen X-3} (V779 Cen) - the stellar parameters are taken from 
Ash et al. (1999) and van der Meer et al. (2007). 
We calculate $P_{ps}/P_{rot}\approx 0.72 \pm0.1$.  
The system is close to equilibrium.
The optical light curve indicates the likely presence of an accretion disc,
but no strong evidence is found for X-ray heating (Tjemkes, van Paradijs \& Zuiderwijk 1986).
The X-ray light curve includes episodes of high and low X-ray flux with a
characteristic timescale of 120-165 d (Priedhorsky \& Terrell 1983; Paul, Raichur \& Mukherjee 2005). 
It could be a beating period between P$_{orb}$ and P$_{rot}$, 
however to prove this we need a higher (better than 1\%) accuracy in 
the measurement of P$_{rot}$.

\subsection{Supergiant systems}

{\bf V830 Cen} (1E 1145.1-6141) - The pulsar appears to be persistent and steady, 
with a typical X-ray flux  of a few mCrab, corresponding to 
a luminosity of about $10^{36}$ erg s$^{-1}$. 
Such a low luminosity is inconsistent with Roche lobe overflow and 
indicates that the pulsar is almost certainly accreting from the wind of V830~Cen. 
The stellar parameters are taken from Ray \& Chakrabarty (2002). 
$P_{ps}/P_{rot}\approx 1.12 \pm 0.15$. The rotation of the mass donor is pseudosynchronized,
the orbit is in a process of circularization. 

{\bf LSI$+65^0010$} (V662 Cas, 2S 0114+650) - the stellar parameters are taken from Grundstrom (2007b).
$P_{ps}/P_{rot}\approx 0.75 \pm 0.1$. The neutron star spins up the
rotation of the mass donor, and the mass donor is close to pseudosynchronization.
No significant X-ray variations are detected in this system. 

{\bf Vela X-1} (GP Vel) -   
$P_{ps}/P_{rot}\approx 0.67 \pm 0.10$. The stellar parameters are 
from Zuiderwijk (1995) and Bildsten et al. (1997). It is  
the brightest persistent accretion powered pulsar. 
The system is close to equilibrium, but the compact object still spins up 
the rotation of the mass donor.

{\bf SMC X-1} (Sk 160) - $P_{ps}/P_{rot}\approx 1.97 \pm 0.07$. The stellar parameters
are taken from van der Meer et al. (2007). SMC X-1 demonstrates an orbital variation of
3.89 days and a superorbital variation with an average length 
of $\sim$ 55 days (Trowbridge, Nowak \& Wilms 2007).
The compact object spins down the rotation of the mass donor.

{\bf BP Cru}  (Wray 977)
The X-ray binary system GX301-2 consists of a neutron star in an eccentric
orbit accreting from the massive early-type star Wray 977. 
The system parameters are from Kaper, van der Meer \& Najarro (2006).
We calculate $P_{ps}/P_{rot}\approx 0.31 \pm 0.03$. The system is
not synchronized  nor circularized.
The tidal force spins up the rotation of the mass donor. 
It has previously been shown that the X-ray orbital light curve is consistent
with the existence of a gas stream flowing out from WRAY 977 in addition to its
strong stellar wind (Leahy \& Kostka 2008).
The stream is (should be) a result of the tidal force.

{\bf Cyg X-1}  (V1357 Cyg) --
With the stellar parameters given in Zi{\'o}{\l}kowski (2005) and Iorio (2008),
we calculate $P_{ps}/P_{rot} \approx 1.04 \pm 0.11$
The system is the brightest persistent source of hard X-rays. 
The Cyg X-1 system is synchronized  and should be circularized, following 
the circularization time in Table 2. 
It is probable that, the measured orbital eccentricity (e=0.06) 
is a spurious effect of the tidal distortion.

\section*{Discussion}
Our goal is to understand, whether the rotation of the mass donors 
in HMXRBs is influenced by the orbiting companion 
(neutron star or stellar mass black hole). 

{\bf Be/X-ray binaries:} In the Be/X-ray binaries the compact object
accretes material from the Be star envelope.
The circumstellar disks around the Be stars in 
Be/X-ray binaries are axisymmetric 
and rotationally supported like the disks in the isolated Be stars; however they 
are smaller and denser (Zamanov et al. 2001).
It seems that transient behaviour in the Be/X-ray binaries
is observed when  the neutron star is located at a distance from the Be star of
$15 < r < 450$ \rsun.
In the Be/X-ray binaries BQ~Cam, V635 Cas, and V725 Tau,  the transient behaviour
can be connected with  the tidal force spinning down the Be star.
Excluding the peculiar object LSI+61$^0$303, for 
these objects typically $P_{ps}/P_{rot} > 10$.

For the galactic microquasar LSI+61$^0$303 the rotation of the mass donor 
is close to pseudosynchronization. The system is known to have been ejected from the cluster IC 1805 
about 1.5 Myr
ago (Mirabel, Rodriguez \& Liu 2004). 
This is the only Be/X-ray binary in which  $\tau_{sync}$ is comparable with the 
life-time of the binary.  

In the long period binary X~Per the neutron star is  far away from the Be star
and the tidal force is weak.

{\bf Giant and Supergiant systems:}
The systems with a giant or supergiant as the mass donor are persistent
X-ray sources and they are close to synchronization/ pseudosynchronization, 
$P_{ps}/P_{rot} \sim 1$. This fact indicates that in these binaries 
the rotation of the mass donors is influenced by the presence 
of the compact object.

In LMC X-4 and Cen X-3, the mass donors (giants) 
are synchronized and the orbits are circularized. 
With respect to the rotation of the mass donor, V725~Tau 
is similar to the Be/X-ray binaries. 

Cyg X-1 is synchronized and almost circularized.  \\
V830~Cen is pseudosynchronized but not circularized yet.
The systems LSI$+65^0010$ and Vela X-1 are close to 
pseudosynchronization and the tidal force accelerates 
the rotation of the mass donors. 
From the calculated $\tau_{sync}$ we can estimate their age 
(the time the neutron star was born) -- 
for  LSI$+65^0010$ \simlt$3.9 \times 10^4$~yr, 
and for V830 Cen \simlt$1.4\times10^4$~yr.
In the case of SMC X-1, the tidal force acts as a decelerator 
of the rotation of the mass donor.
In BP~Cru, a gas stream from the mass donor  exists, 
probably resulting from the strong tidal force and spin-up of the mass donor.

\section*{Conclusions}
In this note we investigate synchronization and
pseudosynchronization in the High-Mass X-ray Binary stars. 
For 13 systems with known orbital and stellar parameters, 
we calculate the synchronization and circularization timescales,
the pseudosynchronization period and compare them
with the data for the rotation of the mass donors.

We find that the Be/X-ray binaries are far away from 
synchronization/pseudosynchronization. For most of them 
$P_{rot} << P_{ps}$. 
The tidal force in the Be/X-ray binaries acts as a decelerator 
of the rotation of the mass donors.  The only Be/X-ray binary
which is close to pseudosynchronization is the peculiar object
LSI$+61^0303$.

The objects containing mass donors of spectral class I (supergiants) 
and III (giants) typically have $P_{rot} \sim P_{ps}$ and are therefore 
close to synchronization /pseudosynchronization.

\acknowledgements
This work was supported in part by  Bulgarian NSF (HTC01-152).
We thank  M. Bode (Liverpool JM University) 
for a throughout reading  of the manuscript,
the referee J. Hurley  for useful comments, 
and A. Gomboc (University of Ljubljana, Slovenia) for the stimulating discussions.


\begin{thebibliography}{}

\bibitem{} Ash, T.~D.~C., Reynolds, A.~P., Roche, P., Norton, A.~J., Still, M.~D., Morales-Rueda, L.: 1999, MNRAS 307, 357 

\bibitem{} Bildsten, L., Chakrabarty, D., Chiu, J., et al.: 1997, ApJS 113, 367 

\bibitem{} Caballero, I., Kretschmar, P., Santangelo, A., et al.: 2007, A\&A 465, L21 

\bibitem{} Campana, S., Stella, L., Mereghetti, S., Colpi, M.: 1995, A\&A 297, 385 

\bibitem{} Casares J., Ribas I., Paredes J.~M., Mart{\'{\i}} J., Allende Prieto C.: 2005, MNRAS 360, 1105

\bibitem{} Claret, A., Gimenez, A.: 1989, A\&AS 81, 37 

\bibitem{} Clark, J.~S., Tarasov, A. E., Steele, I. A., et al.: 1998, MNRAS 294, 165 

\bibitem{} Corbet, R.H.D.: 1984, A\&A 141, 91

\bibitem{} Coe, M.~J., Carpenter, G.~F., Engel, A.~R., Quenby, J.~J.: 1975, Nature 256, 630 

\bibitem{} Coe, M.~J., Reig, P., McBride, V.~A., Galache, J.~L., Fabregat, J.: 2006, MNRAS 368, 447

\bibitem{} Delgado-Mart{\'{\i}}, H., Levine, A.~M., Pfahl, E., Rappaport, S.~A.: 2001, ApJ 546, 455 

\bibitem{} Ducci, L., Sidoli, L., Paizis, A., Mereghetti, S.: 2008, arXiv:0810.5463 

\bibitem{} Germany, L., Proctor, R., Fluke, C., et al.: 2009, 
The Swinburne Astronomy Online Encyclopedia, http://astronomy.swin.edu.au/cosmos/
 
\bibitem{} Grundstrom, E.~D., Blair, J. L., Gies, D. R., et al.: 2007a, ApJ 656, 431

\bibitem{} Grundstrom, E.~D., Boyajian, T. S., Finch, C., et al.: 2007b, ApJ 660, 1398 

\bibitem{} Haigh, N.~J., Coe, M.~J., Fabregat, J.: 2004, MNRAS 350, 1457 

\bibitem{} Hall, D. S.: 1986, ApJ 309, L83

\bibitem{} Heemskerk, M.~H.~M., van Paradijs, J.: 1989, A\&A 223, 154 

\bibitem{} Hurley, J.~R., Tout, C.~A., Pols, O.~R.: 2002, MNRAS 329, 897 

\bibitem{} Hut, P.: 1981, A\&A 99, 126 

\bibitem{} Hutchings, J.~B., Crampton, D.: 1981, PASP 93, 486 

\bibitem{} Iorio, L.: 2008, Ap\&SS 315, 335 

\bibitem{} Kaper, L., van der Meer, A., Najarro, F.: 2006, A\&A 457, 595 

\bibitem{} Kelley, R.~L., Jernigan, J.~G., Levine, A., Petro, L.~D., Rappaport, S.: 1983, ApJ 264, 568 

\bibitem{} Kretschmar, P., Pottsschmidt, K., Ferringo, C., et al.: 2006, in: A. Wilson (ed.), 
{\it The X-ray Universe 2005}, ESA SP-604, p.273 

\bibitem{} Lang, F.~L., Levine, A. M., Bautz, M., et al.: 1981, ApJ 246, L21 

\bibitem{} Leahy, D.~A., Kostka, M.: 2008, MNRAS 384, 747 

\bibitem{} Liu, Q.~Z., van Paradijs, J., van den Heuvel, E.~P.~J.: 2000, A\&AS 147, 25 

\bibitem{} Liu, Q.~Z., van Paradijs, J., van den Heuvel, E.~P.~J.: 2005, A\&A 442, 1135 

\bibitem{} Liu, Q.~Z., van Paradijs, J., van den Heuvel, E.~P.~J.: 2006, A\&A 455, 1165 

\bibitem{} Lyubimkov, L.~S., Rostopchin, S.~I., Roche, P., Tarasov, A.~E.: 1997, MNRAS 286, 549 

\bibitem{} Mirabel, I.F., Rodriguez, L.F., Liu, Q.Z.: 2004, A\&A 422, L29

\bibitem{} Negueruela, I., Okazaki, A.~T., Fabregat, J., Coe, M.~J., Munari, U., Tomov, T.: 2001, A\&A 369, 117 

\bibitem{} Negueruela, I., Okazaki, A.~T.: 2001, A\&A 369, 108 

\bibitem{} Negueruela, I., Roche, P., Fabregat, J., Coe, M.~J.: 1999, MNRAS 307, 695 

\bibitem{} Okazaki, A.~T., Negueruela, I.: 2001, in: R. Giacconi, S. Serio, L. Stella (eds.)
{\it X-ray Astronomy 2000}, ASPC 234, p.281 

\bibitem{} Paul, B., Raichur, H., Mukherjee, U.: 2005, A\&A 442, L15 

\bibitem{} Priedhorsky, W.~C., Terrell, J.: 1983, ApJ 273, 709 

\bibitem{} Ray, P.~S.,  Chakrabarty, D.: 2002, ApJ 581, 1293 

\bibitem{} Reig, P., Larionov, V., Negueruela, I., Arkharov, A.~A., Kudryavtseva, N.~A.: 2007, A\&A 462, 1081 

\bibitem{} Roche, P., Larionov, V., Tarasov, A. E., et al.: 1997, A\&A 322, 139 

\bibitem{} Rosenberg, F.~D., Eyles, C.~J., Skinner, G.~K., Willmore, A.~P.: 1975, Nature 256, 628 

\bibitem{} Smith, D.~M., Hazelton, B., Coburn, W., et al.: 2005, ATEL 557, 1 

\bibitem{} Tjemkes, S.~A., van Paradijs, J., Zuiderwijk, E.~J.: 1986, A\&A 154, 77 

\bibitem{} Trowbridge, S., Nowak, M.~A., Wilms, J.: 2007, ApJ 670, 624 

\bibitem{} Tueller, J., Ajello, M., Barthelmy, S., Krimm, H., Makwardt, C., Skinner, G.: 2005, ATEL 504, 1 

\bibitem{} van der Meer, A., Kaper, L., van Kerkwijk, M.~H., Heemskerk, M.~H.~M., van den Heuvel, E.~P.~J.: 2007, A\&A 473, 523 

\bibitem{} Zahn, J.-P.: 1975, A\&A 41, 329 

\bibitem{} Zamanov, R.~K., Reig, P., Mart{\'{\i}}, J., Coe, M.~J., Fabregat, J., Tomov, N.~A., Valchev, T.: 2001, A\&A 367, 884 

\bibitem{} Zi{\'o}{\l}kowski, J.: 2005, ChJAS 5, 75 

\bibitem{} Zuiderwijk, E.~J.: 1995, A\&A 299, 79 

\end{thebibliography}
\end{document}